\documentclass[12pt]{article}
\title{    Shadows  and Twisted Variables  }
\usepackage{mathrsfs}
\usepackage{amsmath}

\global\arraycolsep=1pt
\usepackage[colorlinks=true,backref=true,linkcolor=black,anchorcolor=black,citecolor=black,filecolor=black,menucolor=black,pagecolor=black,urlcolor=black]{hyperref}
\usepackage[Symbol]{upgreek}
\usepackage{bbm}
\usepackage{dsfont}
\usepackage{amssymb}
\usepackage{textcomp}
 
\newcommand{\baa}{/ \hspace{-1.4ex}}
\newcommand{\baaa}{\, / \hspace{-1.6ex}}

\newcommand{\Scal}[1]{\biggl ({#1} \biggr )}
\newcommand{\scal}[1]{\bigl ({#1} \bigr )}
\def\bea{\begin{eqnarray}}
\def\eea{\end{eqnarray}}
\def\be{\begin{equation}}
\def\ee{\end{equation}}
\newcommand{\CR}{\nonumber \\*}

\newcommand{\trace}{\hbox {Tr}~}

\newcommand{\gra}[2]{{\scriptscriptstyle (#1 , #2 )}}
\newcommand{\ur}[1]{{\scriptscriptstyle (#1)}}
\def\L{{\cal L}}

\DeclareMathAlphabet{\mathpzc}{OT1}{pzc}{m}{it}
\def\s{\,\mathpzc{s}\,}
\def\a{{\scriptscriptstyle (\mathpzc{s})}}
\def\q{{{\scriptscriptstyle (Q)}}}
\def\qs{{\scriptscriptstyle (Q\mathpzc{s})}}

\def\Lc{\mathscr{L}}

\def\Q{{\mathcal{S}_{\q}}}
\def\hQ{{\mathcal{S}^{ \scriptscriptstyle \rm ext}_{\q}}}
\newcommand\phis[1]{{\phi^\a_{#1}}}
\newcommand\phiq[1]{{\phi^{\q}_{#1}}}
\newcommand\phiqs[1]{{\phi^{{\qs}}_{#1}}}
\def\blambdas{{\overline{\lambda}^\a}}
\def\blambdaq{{\overline{\lambda}^{\q}}}
\def\blambdaqs{{\overline{\lambda}^{{\qs}}}}
\def\lambdaq{{\lambda^{\q}}}
\def\lambdaqs{{\lambda^{{\qs}}}}
\def\Omegas{{\Omega^\a}}
\def\Omegaq{\Omega^{\q}}
\def\Omegaqs{\Omega^{{\qs}}}
\newcommand\As[1]{{A^\a_{#1}}}
\newcommand\Aq[1]{A^{\q}_{#1}}
\newcommand\Aqs[1]{A^{{\qs}}_{#1}}
\def\cq{c^{{\q}}}
\def\muq{\mu^{{\q}}}

\def\C{c}

\def\S{{\mathcal{S}_\a}}

\def\F{\mathscr{F}}

\newcommand{\ins}[1]{\S_{|\Gamma} \bigl[
\hat \Uppsi^{\scriptscriptstyle{(#1)}} \cdot \Gamma \bigr]}

\usepackage[colorlinks=true,backref=true,linkcolor=black,anchorcolor=black,citecolor=black,filecolor=black,menucolor=black,pagecolor=black,urlcolor=black]{hyperref}

\usepackage[Symbol]{upgreek}
\usepackage{caption}
\usepackage{bbm}
\usepackage{dsfont}
\usepackage{amssymb}
\usepackage{textcomp}
\def\bea{\begin{eqnarray}}
\def\eea{\end{eqnarray}}
\def\be{\begin{equation}}
\def\ee{\end{equation}}

\def\L{{\cal L}}

\def\Lc{\mathscr{L}}

\newcommand\vo[1]{{\, ^{{ \scriptscriptstyle #1}} \hspace{-0.8mm} v}}
\def\k{{\rm\scriptscriptstyle K}}
\def\bps{{\rm \scriptscriptstyle C}}
\def\uk{{\, ^\k \hspace{-0.7mm} u}}
\def\ub{{\, ^\bps \hspace{-0.7mm} u}}

\def\susy{{\delta^{\mathpzc{Susy}}}}

\def\N{\mathcal{N}}
\begin{document}
\allowdisplaybreaks[1]
\renewcommand{\thefootnote}{\fnsymbol{footnote}}
\def\corr{$\spadesuit $}
\def\trefle{ $\clubsuit$}
\begin{titlepage}
%\null
\begin{flushright}
CERN-PH-TH/2007-009
% hep-th/
\end{flushright}
\vskip 3em
\begin{center}
{{\Large \bf
{ Shadows  and Twisted Variables }
}}
\lineskip .75em
\vskip 3em
\normalsize
{\large Laurent Baulieu\footnote{email address: baulieu@lpthe.jussieu.fr} and 
Guillaume Bossard\footnote{email address: bossard@lpthe.jussieu.fr}\\
$^{* }$\it Theoretical Division CERN, \\{ CH-1211 Geneva  23, Switzerland
}
\\
$^{*\dagger}$ {\it Universit\'e   Pierre et Marie Curie, LPTHE, CNRS  \\
4 place Jussieu, F-75252 Paris Cedex 05, France}
\\
$^{\dagger}${\it Instituut voor Theoretische Fysica\\
Valckenierstraat 65, 1018XE Amsterdam, The Netherlands}}

\vskip 3em
\vskip 1 em
\end{center}
\vskip 1 em
\begin{abstract}
\end{abstract}
We explain how a new type of fields called shadows and the use of twisted variables allow for a better  description of  Yang--Mills supersymmetric theories.

 \noindent
 (Based on lectures given in Carg\`ese, June 2006.)

%Using the shadow dependent  decoupled Slavnov--Taylor identities associated to gauge invariance and supersymmetry, 
%%that are allowed by the introduction of shadow fields, 
%we discuss the  renormalization of   the  $\N=4$ super-Yang--Mills theory and of  its coupling to gauge-invariant  operators. We specify the method for    the determination of non-supersymmetric counterterms that are needed to maintain supersymmetry.
% in a way that preserves supersymmetry. 
%This method justifies the previous work of St\"{o}ckinger et al, in the case of $\N=1$.
\end{titlepage}
%%%%%%%%%%%%%%%%%
\renewcommand{\thefootnote}{\arabic{footnote}}
\setcounter{footnote}{0}

%%%%%%%%%%%%%%%%%%%%%%%%%%%%%%%%%%%%%%%%%%%%%%%%%%%%%%%%

%%%%%%%%%%%%%%%%%%%%%%%%%%%%%%%%%%%%%%%%%%%%%%%%%%%%

\renewcommand{\thefootnote}{\arabic{footnote}}
\setcounter{footnote}{0}

%%%%%%%%%%%%%%%%%%%%%%%%%%%%%%%%%%%%%%%%%%%%%%%%%%%%%%%%

%\tableofcontents
%%%%%%%%%%%%%%%%%%%%%%%%%%%%%%%%%%%%%%%%%%%%%%%%%%%%%%%%

\section{Introduction}

Non-linear aspects   and   the non-existence  of a supersymmetry-preserving regulator make the definition of  supersymmetric theories a subtle task.  We explain in these lectures notes  that  the introduction of new fields, called shadows, clarify the construction of Yang--Mills supersymmetric theories.

In the formalism that we develop,  a  supersymmetric theory is   defined  in terms  of classical fields (gauge fields and matter fields), Faddeev--Popov ghosts and shadow fields.  
 Gauge invariance is expressed   by  the   BRST invariance, with a   graded differential  operator~$\s$. 
The  shadows    fields    permit  the replacement of the notion of    the  supersymmetry generators  by  that of a  differential  operator $Q$,  consistent with $\s$.
The operator $Q$ acts as an ordinary  supersymmetry transformation on the gauge invariant functions of the physical fields.  Moreover, there exist gauges for which $Q$ annihilates  both the classical action and the    $\s$-exact gauge-fixing action.

The advantage of having  both  operators  $\s$ and  $Q$ acting on   the extended set of  fields   is that two independent Slavnov--Taylor identities can be associated with    supersymmetry and BRST invariances.
 Observables  can be      appropriately defined  for understanding their gauge and supersymmetry covariance : they are the cohomology of the BRST symmetry. Anomalies and renormalization can be conventionally analyzed, considering insertions of arbitrary composite operators. 
  This defines  an  unambiguous   renormalization  process of    Yang--Mills   supersymmetric theory, for any given choice of the regularization of divergences.

Shadows can be used   to demonstrate non-renormalization theorems. Moreover, the  proofs are greatly simplified by   twisting the spinor fields in tensors.  In fact, twisted variables permit one to determine  off-shell closed   sub-sectors  of   supersymmetry algebra that are  relevant for the non-renormalization properties.

Both differential operators $\s$ and $Q$ of  supersymmetric  theories  satisfy extended curvature conditions, analogous to those   of the   topological BRST operator  of topological quantum field theory. This similarity suggests   that    some of the relevant  equations   for the non-renormalization theorems have a    geometrical  meaning . 

\section{Introducing the shadow fields}

To fix ideas, consider the $\N=4,\ d=4$ supersymmetric action in flat space. The physical fields of   this gauge invariant theory with $SO(3,1)$ Lorentz symmetry are the gauge field
$A_\mu$, the $SU(4)$-Majorana spinor $\lambda$,  and the six scalar fields
$\phi^i$ in the vector representation of $SO(6) \sim SU(4)$. All fields   are
  in the adjoint representation of a compact gauge group that we
will suppose simple. The classical action is uniquely determined by  supersymmetry,  $Spin(3,1)\times SU(4)$ global symmetry  and gauge invariance. It reads
\be S \equiv \int d^4 x \trace \Bigl(- \frac{1}{4} F_{\mu\nu} F^{\mu\nu} -
\frac{1}{2} D_\mu \phi^i D^\mu \phi_i + \frac{i}{2}
\scal{\overline{\lambda} \baaa D \lambda } - \frac{1}{2} \scal{
\overline{\lambda} [ \phi, \lambda]} - \frac{1}{4}
[\phi^i,\phi^j][\phi_i,\phi_j] \Bigr) \ee
with $\phi \equiv \phi^i \tau_i $ and the supersymmetry transformations $\susy$  
\begin{gather}
\susy A_\mu = i \scal{\overline{\epsilon} \gamma_\mu \lambda }
\hspace{10mm} \susy \phi^i = - \scal{ \overline{\epsilon} \tau^i
\lambda}
\hspace{10mm} 
% \CR
\susy \lambda = \scal{ \baa F + i \baaa D \phi + \frac{1}{2} [\phi,
\phi]} \epsilon 
\end{gather}

For the sake of convenience,  we can chose the parameter $\epsilon$ as  a commuting spinor.  In this way,  $ \susy^2$ represents the commutator of two supersymmetry transformations, with
 \bea \label{bclosure}\susy^2\approx \delta^{\rm gauge}(\overline{\epsilon} [ \phi - i
\baaa A ] \epsilon) - i (\overline{\epsilon} \gamma^\mu \epsilon)
\partial_\mu\eea
Here $\approx $ stands for the equality modulo equations of motion.

In view of   the last equation, the   quest of a quantum field theory with supersymmetry   implies  the following remarks.

The presence   of   equations of motion in the right-hand-side of   (\ref{bclosure})   is a rather   annoying technical difficulty. However, it  can  always be   turned around in quantum field theory, by using the Batalin-Vilkowiski formalism. Moreover, as we will shortly see, even in the case where no auxiliary fields exist,  it can be practically  resolved  in the  proofs for the consistency of the quantum theory by  using twisted variables.

   The existence of the  field dependent gauge transformation in the commutator of two supersymmetry transformations   (\ref{bclosure}) is a deeper problems.  
%   It forbids the interpretation of   $\susy$ as the  "infinitesimal"  transformations of a well-defined symmetry acting on local fields  with a group structure.   
    It  concretely  implies that   one cannot give   sense to the notion of a $\susy$-invariant     gauge-fixing action.  This fact explicitly shows up when one uses  the Faddeev-Popov procedure. Suppose that 
 one fixes the gauge, say in a Feynman--Landau gauge. This process is independent of supersymmetry and gives an action
\be S _{\rm gf} =S + \int \trace  \Bigl( {(\partial A )^2\over{2 \alpha}} -\bar \Omega \partial  D  \Omega \Bigr)
\ee
This lagrangian    breaks gauge invariance in the desired  way, but  one cannot find a  definition of $\susy$ acting on the scalar Faddeev--Popov ghosts $\Omega$ and $\bar\Omega$ that is compatible with the closure relation (\ref{bclosure}).      This forbids one to  define the Ward identities associated to supersymmetry  with    usual techniques.  Therefore, one must  improve the techniques  currently used for ordinary global symmetries coupled to gauge invariance. Since  there are cases where an off-shell  superfield formalism  does not exist (in particular for the $\N=4$ theory) and since no regulator exist that can maintain both supersymmetry and gauge invariance, such improvement must follow  from  new ideas.  

One  method for handling the  problems caused by    the  gauge transformations in the  closing relations for the   supersymmetry transformations of classical fields    is by  introducing an additional anticommuting scalar field $c$    valued in the Lie algebra  of the gauge group.  On can define in this way a differential  operator $Q$  out of $\susy$,  which is nilpotent modulo a translation \cite{shadow}
 \be \label{Qclosure}Q^2\approx  - i (\overline{\epsilon} \gamma^\mu \epsilon)
\partial_\mu\ee
The way to do so is to define   the action of  $Q$    on all the physical fields   $\varphi$  and $c$ as follows 
 \be Q  \varphi=\susy ({\epsilon})  \varphi-\delta^{\rm gauge}(c) \varphi
 \CR
 \ee
with
 \be \label{defQ}
Q c = ( \overline{\epsilon} [ \phi - i
\baaa A ] \epsilon) - c^2 
\ee

The field $c$ will be called the shadow field, and its presence will allow one to solve at once all questions discussed above, with the conclusion  that the notion of the operator $\susy$ must be replaced by that  of the differential $Q$ at the quantum level, in a way that is analogous to the  enhancement of  gauge invariance into     BRST symmetry.

We see that the action of $Q$   on the classical fields is linear in the global  parameters  ${\epsilon}$ and  on the field $c$. Since, for the classical fields,  $Q$ is the sum of a supersymmetry transformation and a gauge transformation, $\susy$ invariance is the same as $Q$ invariance for gauge invariant quantities. 

The action of $Q$ on $c$ is quadratic  both in $c$ and $\epsilon$, and $Q \epsilon=0$. We   have the existence of a grading equal to the shadow number, which is zero for the classical fields, and one  for $c$ and $\epsilon$.

 In practice,   one must do computations with a BRST invariant gauge-fixed theory, where interacting   Faddeev--Popov  ghosts propagate. In fact, renormalization generally  mixes gauge invariant operators  with non gauge-invariant    BRST-exact operators. Thus, observables must be defined  through the  cohomology of the BRST operator $\s$ for ordinary gauge symmetry.  To control the  covariance under supersymmetry of observables,  the BRST Ward identity and the   supersymmetry  Ward identities  must be  disentangled. It follows that $Q$ and $\s$ must be independent and consistent  operators (i.e., $Q$  and $\s$ must anticommute). Therefore the scalar field $c$ cannot be identified with the Faddeev--Popov ghost $\Omega$.  
  
 The idea of shadows \cite{shadow} is thus  to introduce new fields, in the form of BRST doublets, in order not to affect physical quantities, and to   redefine  the supersymmetry transformations  of classical fields  by addition of a  compensating gauge transformations with a parameter equal to the   shadow field $c$.  Moreover,  Eq.~(\ref{Qclosure})  must be  satisfied for all fields.

The action of the BRST operator $\s $  on all physical fields is nothing  but a gauge transformation of parameter $\Omega$  with
\be
\s\varphi  =    - \delta^{\rm gauge}( \Omega)  \varphi\hspace{10mm}
\s \Omega =-\Omega^2
\ee
and since the shadow $c$ must not affect the physical sector of the theory we introduce the commuting scalar $\mu$ such that $(c,\mu)$      builds a trivial BRST doublet
\be \s c = \mu \hspace{10mm} \s \mu
= 0 \ee
We want to impose   Eq.~(\ref{Qclosure})  on all fields, as well as
\be
\label{sclosure}
\s^2=\s Q+Q\s=0
\ee
In fact, by a direct computation, we find that the algebra (\ref{Qclosure}) and (\ref{sclosure}) is satisfied with 
 \be
Q \Omega = - \mu - [c, \Omega] 
\hspace{7mm} 
   Q \mu = - [ (
\overline{\epsilon} \phi \epsilon ) , \Omega ] + i (\overline{\epsilon}
\gamma^\mu \epsilon) D_\mu \Omega - [c, \mu]
\ee
We will  shortly write a curvature   equation that explains these transformation laws, and in particular the  property
\be
\s c   +Q \Omega  + [c, \Omega] = 0
\ee

In order to define the Ward identities associated to supersymmetry, we need a BRST-exact gauge-fixing that is $Q$-invariant. Such gauge-fixing will be said to be supersymmetric. To define it, we introduce the trivial quartet $\bar \mu ,\, \bar c,\,\bar \Omega,\, b$, with
\begin{gather}\begin{split}
\s \bar \mu &= \bar c \\* Q \bar \mu &= \bar \Omega
\end{split}\hspace{10mm}\begin{split}
\s \bar c &= 0 \\* Q \bar c &= - b
\end{split}\hspace{10mm}\begin{split}
\s \bar \Omega &= b \\ Q \bar \Omega &= - i (\overline{\epsilon}
\gamma^\mu \epsilon) \partial_\mu \bar \mu
\end{split}\hspace{10mm}\begin{split}
\s b &=0 \\* Q b&= i (\overline{\epsilon}
\gamma^\mu \epsilon) \partial_\mu \bar c
\end{split}\end{gather}
%On all fields, we thus  have $
%\s^2 = 0,\ Q^2 \approx -i (\overline{\epsilon} \gamma^\mu
%\epsilon) \partial_\mu,\ 
%\{\s , Q \} = 0
%$. 

%The shadow
%fields have been assembled into BRST doublets, so that  they  do not affect the physical sector. 
The quantum field theory has an internal 
bigrading, the ordinary ghost number and the new shadow
number.    The $Q$ transformation of fields   depend on the constant commuting supersymmetry parameter.  The latter  is understood as an ordinary gauge parameter for the quantum field theory,
but observables will not depend on them, owing to BRST invariance.

%From now on we see that we solved the questions  in the remarks 1) and 2).

%Having both Slavnov--Taylor identities for gauge invariance
%and supersymmetry is a safe starting point to define
%perturbative theory. 

\section {Supersymmetric shadow dependent lagrangians}

In order to control supersymmetry and  renormalize the theory,  we start  from a renormalizable $\s$ and $Q$ invariant gauge-fixed action, which determines the Feynman rules. A class of such actions is of the form:
\be \label{SLG}
S_{\rm gf} [\varphi, \Omega,\bar \Omega, b, c, \bar  c, \mu,\bar\mu]=
S[\varphi] - \s Q \int \trace \bar\mu \Bigl(\partial A +{\alpha  \over 2} b\Bigr)
\ee
One has indeed
\bea
- \s Q \int \trace \bar\mu \Bigl(\partial A +{\alpha  \over 2} b\Bigr)
=-\s \int  \trace \Bigl( \bar \Omega \Bigl(\partial A +{\alpha  \over 2} b\Bigr) +\bar \mu Q \Bigl( \partial A +{\alpha  \over 2} b\Bigr)\Bigr)\CR
=\int \trace \Bigl(-{\alpha  \over 2} b^2  -  b\partial A  - \bar \Omega \partial D \Omega +\ldots\Bigr)
\eea
Here,   the dots  stand for terms  that imply a propagation of the pairs of shadows $\mu,\bar \mu$ and 
$c,\bar c$.   They are  given by an easy computation.  They imply    $\epsilon$-dependent propagators and vertices.  However, observables are defined by the cohomology of the BRST operator $\s$, so that their expectation values are independent on the values of $\epsilon$, since the later occur through an $\s$-exact term. 

In the absence of anomaly, one can enforce both Ward identities for the $\s$ and $Q$ invariances. This  means that one can concretely impose renormalization  conditions which enforce these identities at any given finite order of perturbation theory, within the framework of   any type of regularization for divergences.

The  prize one has to pay for having shadows is that they generate   a perturbative theory  with  more   Feynman diagrams. If we  consider physical composite operators
that mix through renormalization with BRST-exact operators, the latter can depend on all possible fields that propagate, and we  have  in
principle to consider a dependence on the  whole set of fields in order to compute the
supersymmetry-restoring non-invariant counterterms. For certain  ``simple"  Green
functions, which cannot mix  with  BRST-exact composite
operators, there exist gauges  in which some of the additional fields can be
integrated out, in a way that justifies, a posteriori, the work of St\"{o}ckinger et
al. for  the $\N=1$ theories \cite{stockinger}. By doing this elimination, one   loses the algebraic
meaning, but one   may gain in computational simplicity.

%
%.\footnote{We do not
%exclude the possibility   of  also    reducing  the set of fields in
%the general case, including observables that mix with BRST-exact
%operators through renormalization, but further investigations are
%needed in order to establish this statement. }
%From a technical point of view, for the class of gauges used by St\"{o}ckinger et al., the use of a single Slavnov--Taylor identity is sufficient, because the square of two supersymmetry transformations on classical fields contains a gauge transformation, while there are equations of motions that can been enforced as Ward identities, and determine the ghost dependence of the
% 1PI generating functional. However the field that was understood as the Faddeev--Popov ghost is in fact the shadow, whereas the true Faddeev--Popov ghost decouples from the theory for correlation functions that do not include it. Moreover, 

%For extended supersymmetry such as the $\N =4$ super-Yang--Mills theory, more care has to be done,  in order to maintain the $R$ symmetry and the supersymmetry . 

%and  observables
%are usually defined as   correlation functions of gauge-invariant  operators
%
%Recent practical computations using dimensional reduction have
%apparently produced inconsistencies in their three loop quantum
%computations. They might be related to a violation of supersymmetric
%Ward identities.
  The  shadow dependent  methodology    is suitable  for  non-ambiguously  computing   the non-invariant counterterms that  maintain supersymmetry, BRST invariance  and the R-symmetry. It applies  to  the renormalization of   all  supersymmetric theories.

\section{Renormalization}

  \subsection{Ward identities for the theory}
By introducing  sources associated to the non-linear  $\s$, $Q$  and $\s Q$ transformations of fields, we  get the following   $\epsilon$-dependent action, which   initiates a BRST-invariant  supersymmetric perturbation theory\footnote{
$M$ is the $32 \times 32$ matrix 
$ M \equiv \frac{1}{2} (\overline{\epsilon} \gamma^\mu \epsilon)
\gamma_\mu + \frac{1}{2} (\overline{\epsilon} \tau_i \epsilon)
\tau^i - \epsilon \overline{\epsilon} $. It occurs 
 because $Q^2$ is a pure derivative only modulo
 equations of motion.
 The 
dimension of $A_\mu,\, \lambda ,\, \phi^i,\, \Omega,\, \bar \Omega ,\,
b,\, \mu,\, \bar \mu,\, c$ and $\bar c$ are respectively $1,\,
\frac{3}{2},\, 1,\, 0,\, 2,\, 2,\, \frac{1}{2},\, \frac{3}{2},\,
\frac{1}{2}$ and $\frac{3}{2}$. Their ghost and shadow numbers are respectively
$(0,0),\,(0,0),\,(0,0),\,
(1,0),\,(-1,0),\,(0,0),\,(1,1),\,(-1,-1),\,(0,1) $ and $(0,-1)$.}
\begin{multline} 
\Sigma \equiv \frac{1}{g^2} S - \int d^4 x \trace \Bigl( b
\partial^\mu A_\mu + \frac{\alpha}{2} b^2 - \bar c
\partial^\mu \scal{D_\mu c + i(\overline{\epsilon} \gamma_\mu \lambda)}
- \frac{i \alpha}{2} (\overline{\epsilon} \gamma^\mu \epsilon) \bar c
\partial_\mu \bar c \\*\hspace{50mm} + \bar \Omega \partial^\mu
D_\mu \Omega - \bar \mu \partial^\mu \scal{D_\mu \mu + [D_\mu \Omega, c] - i
(\overline{\epsilon} \gamma_\mu [\Omega, \lambda])} \Bigr) \\*
+ \int d^4 x \trace \biggl( \As{\mu} D^\mu \Omega + \blambdas [\Omega,
\lambda] - \phis{i} [\Omega , \phi^i] + \Aq{\mu} Q A^\mu - \blambdaq Q \lambda + \phiq{i} Q \phi^i \\* 
+ \Aqs{\mu} \s Q A^\mu - \blambdaqs \s Q \lambda + \phiqs{i} \s Q
\phi^i + \Omegas \Omega^2 -
\Omegaq Q \Omega - \Omegaqs \s Q \Omega \\* - \cq Q c + \muq Q \mu
+ \frac{\, g^2}{2} ( \blambdaq - [\blambdaqs , \Omega ]) M ( \lambdaq -
[\lambdaqs , \Omega ]) \biggr)
\end{multline}

Because of the $\s$ and $Q$ invariances, the action is invariant under the both 
Slavnov--Taylor identities defined in \cite{shadow}, which are  associated
respectively to gauge and supersymmetry invariance,
$ \S(\Sigma) = \Q(\Sigma) = 0$. For the sake of illustration, let  us present the  supersymmetry Slavnov--Taylor operator of the $\N=4 $ theory\footnote{ The linearized Slavnov--Taylor operator  ${\Q _{|\Sigma}}$   \cite{shadow} verifies 
  ${\Q _{|\Sigma}} ^2 =-i (\overline{\epsilon} \gamma^\mu
\epsilon) \partial_\mu$, which solves in practice the fact that $Q^2$
is a pure derivative only modulo equations of motion.}
\begin{multline}
\Q(\F) \equiv \int d^4 x \trace \biggl( \frac{\delta^R \F}{\delta A^\mu}
\frac{\delta^L \F}{\delta \Aq{\mu}} + \frac{\delta^R \F}{\delta \lambda}
\frac{\delta^L \F}{\delta \blambdaq}+ \frac{\delta^R \F}{\delta \phi^i}
\frac{\delta^L \F}{\delta \phiq{i}} + \frac{\delta^R \F}{\delta c}
\frac{\delta^L \F}{\delta \cq} +\frac{\delta^R \F}{\delta
\mu}\frac{\delta^L \F}{\delta \muq} \biggr .\\* + \frac{\delta^R \F}{\delta \Omega}
\frac{\delta^L \F}{\delta \Omegaq} - \As{\mu}
\frac{\delta^L \F}{\delta \Aqs{\mu}} + \blambdas \frac{\delta^L \F}{\delta
\blambdaqs} - \phis{i} \frac{\delta^L \F}{\delta          
\phiqs{i}} + \Omegas \frac{\delta^L \F}{\delta \Omegaqs} - b
\frac{\delta^L \F}{\delta \bar c} + \bar
\Omega \frac{\delta^L \F}{\delta \bar \mu} 
\\*-i (\overline{\epsilon} \gamma^\mu \epsilon)\Bigl(- \partial_\mu
\Aqs{\nu} \frac{\delta^L \F}{\delta \As{\nu}} + \partial_\mu
\blambdaqs \frac{\delta^L \F}{\delta \blambdas}- \partial_\mu
\phiqs{i} \frac{\delta^L \F}{\delta \phis{i}} + \partial_\mu
\Omegaqs \frac{\delta^L \F}{\delta \Omegas} - \partial_\mu \bar c \frac{\delta^L \F}{\delta b} + \partial_\mu \bar \mu \frac{\delta^L \F}{\delta \bar \Omega} \Bigr . \\* + \Aq{\nu} \partial_\mu 
A^\nu + \blambdaq \partial_\mu \lambda + \phiq{i} \partial_\mu
\phi^i + \Omegaq
\partial_\mu \Omega+ \cq \partial_\mu c + \muq \partial_\mu \mu \Bigr) \biggr)
\label{slavnovQ}
\end{multline}
If no anomaly occurs,  the  Slavnov--Taylor  identities  $ \S(\Gamma) = \Q(\Gamma) = 0$ completely determines   all ambiguities of the  supersymmetric effective action $\Gamma$, order by order in perturbation theory.

\subsection{Anomalies}
In \cite{shadow,beta}, we   showed the absence of anomaly for the  $\N=2,4$ and the
stability of the $\N=1,2,4$ action $\Sigma$ under renormalization. Thus, all   Green functions of 
the complete theory involving shadows and ghosts can be renormalized,
in any given regularization scheme, so that supersymmetry and gauge
invariance are preserved at any given finite order.

Let us   sketch the proof that no supersymmetry anomaly can exist for $\N=2,4$, and that for $\N=1$ the only possible anomaly is the Adler--Bardeen anomaly.  

An anomaly in a supersymmetry theory can only occur if a pair of  local functionals  $\cal A$ and $\cal B$  of the fields  and sources
can violate the pair of Ward identities for both $\s$ and $Q$ invariances. For instance, when one renormalizes  the theory at the one-loop level,  the result of the computation can violate the  Ward identities 
by   local terms  $\cal A$ and $\cal B$,   as follows
\bea\label{ano}
 { \S}_{|{\Sigma}}  \Gamma^{\rm 1\ loop}= \hbar\int  {  \cal A   }\hspace{10mm}  { \Q}_{|{\Sigma}} \Gamma^{\rm 1\ loop}= \hbar\int  {  \cal B   }
\eea
If either    $\cal A$ and $\cal B$  cannot be eliminated by adding local counterterms to   $\Gamma^{\rm 1\ loop}$, which means that they are not
$ { \S}_{|{\Sigma}}$ and  ${ \Q}_{|{\Sigma}}$ exact,  
 one has an anomaly, and the theory cannot be renormalized while maintaining either supersymmetry or gauge invariance, or both. In \cite{shadow,beta}, we proved that  the solution $\cal A$ and $\cal B$   of Eq.~(\ref{ano}), modulo $ { \S}_{|{\Sigma}}$ and  ${ \Q}_{|{\Sigma}}$  exact terms,  can only depend on the fields, and thus, the consistency relation for $\s$ and $Q$ implies:
\be
\s \int  {  \cal A   }  = 0
\ \ \ \ \ 
Q \int     {  \cal A   }+\s  \int {    \cal B   }=0
\ \ \ \ \ \
Q \int   {    \cal B   }= 0
\ee
In fact, the  first equation implies that  $\cal A$ must be  the  consistent Adler-Bardeen  anomaly, which descends formally from  the   Chern class $\trace FFF$. But then, the $Q$ symmetry is so demanding that  the second and third equations have  no  solution  $ {  \cal B   }\neq 0$ for  $\N=2,4$.  Thus there  cannot be an anomaly for these cases. For  $\N=1$, the constraint is weaker, and the Adler-Bardeen  anomaly admits a supersymmetric counterpart $ {  \cal B   }$. However, the Adler--Bardeen theorem holds, and if the one-loop coefficient of the Adler--Bardeen anomaly cancels, it will cancel to all order.

Of course, these are  well known facts.  However,  by having introduced the shadows,  both Ward identities for supersymmetry and gauge invariance allow a   safe verification of the status of gauge and supersymmetry anomalies by the standard consistency argument, valid to all order of perturbation theory.

\subsection{ Ward identities for the observables}
Observables of   a   super-Yang--Mills theory  are Green functions of
local operators in the cohomology of the BRST linearized Slavnov--Taylor operator $\S_{|\Sigma}$.  From this definition,  these      Green functions   are independent of the gauge parameters of the action, including~$\epsilon$.  Classically, they are represented by  gauge-invariant  polynomials of the physical fields \cite{shadow,henneaux}. We  introduce classical sources $u$ for all these operators.  We must generalize the supersymmetry Slavnov--Taylor identity for the extended local action that   depends on  these sources. Since the supersymmetry algebra does not close off-shell,   other sources $v$, coupled to unphysical $\S_{|\Sigma}$-exact operators, must also be introduced. We define the following field and source combinations $\varphi^*$
\be\begin{split}
A^*_\mu &\equiv \Aq{\mu} - \partial_\mu \bar c - [\Aqs{\mu} - \partial_\mu \bar \mu , \Omega] \\*
\phi^*_i &\equiv \phiq{i} - [\phiqs{i} , \Omega]
\end{split}\hspace{10mm}\begin{split}
c^* &\equiv \cq - [\muq,\Omega] \\*
\lambda^* &\equiv \lambdaq - [\lambdaqs , \Omega] 
\end{split}\ee
They verify $\S_{|\Sigma} \varphi^* = - [\Omega, \varphi^*]$. The  collection  of local operators coupled to the $v$'s  is made of all possible gauge-invariant (i.e. $\S_{|\Sigma}$-invariant) 
 polynomials in the physical fields and   the $\varphi^*$'s. These operators have ghost number zero, and their shadow number is negative, in contrast with the physical gauge-invariant  operators, which have shadow number zero.

The relevant action is thus $\Sigma[u,v] \equiv \Sigma + \Upsilon[u,v]$, with 
\begin{multline}
\Upsilon[u,v] \equiv \int d^4 x \biggl( u_{ij} \frac{1}{2}
\trace \phi^i \phi^j + u^\alpha_i \trace \phi^i \lambda_\alpha +
u_{ijk} \frac{1}{3}\trace \phi^i \phi^j \phi^k \\*+ \uk_{ij}^\mu \trace \scal{ i \phi^{[i}
D_\mu \phi^{j]} + \frac{1}{8} \overline{\lambda} \gamma_\mu \tau^{ij}
\lambda} + \uk^{\mu\nu}_i \trace \scal{ F_{\mu\nu} \phi^i
-\frac{1}{2} \overline{\lambda} \gamma_{\mu\nu} \tau^i \lambda} + \uk^5_\mu \frac{1}{2} \trace \overline{\lambda}
\gamma_5 \gamma^\mu \lambda \\*+ \ub_{ijk} \trace\scal{\frac{1}{3}
 \phi^{[i} \phi^j \phi^{k]} + \frac{1}{8} \overline{\lambda} \tau^{ijk}
 \lambda } + \ub_{ij}^\mu \trace \scal{ i \phi^{[i}
D_\mu \phi^{j]} - \frac{1}{4} \overline{\lambda} \gamma_\mu \tau^{ij}
\lambda} \\* + \ub^{\mu\nu}_i \trace \scal{ F_{\mu\nu} \phi^i +\frac{1}{4}
\overline{\lambda} \gamma_{\mu\nu} \tau^i \lambda} + u_{ij}^\alpha
\trace \phi^i \phi^j \lambda_\alpha + i u_i^{\mu\, \alpha} \trace D_\mu
\phi^i \lambda_\alpha + u^{\mu\nu\, \alpha} \trace F_{\mu\nu}
\lambda_\alpha + \cdots \\*
+ v_i^\alpha \trace
\phi^i \lambda^*_\alpha + v^{\alpha\beta} \trace \lambda_\alpha
\lambda^*_\beta + v_i^\mu \trace \phi^i A^*_\mu + v_{ij} \trace \phi^i
\phi^{*\, j} + i v^{\mu\, \alpha}_i \trace D_\mu \phi^i \lambda^*_\alpha \hspace{9mm}\\*+
\vo{0}^\alpha_i \trace \lambda_\alpha \phi^{*\, i} + i v^{\mu
 \alpha\beta} \trace D_\mu \lambda_\alpha \lambda^*_\beta + i 
v^\mu_{ij} \trace D_\mu \phi^i \phi^{*\, j} + i \vo{-1}_i^{\mu\alpha} \trace D_\mu
\lambda_\alpha \phi^{*\, i} + \cdots \biggr)
\end{multline} 
Here, the  $\cdots$ stand for all   other analogous  operators.

The Slavnov--Taylor operator $\Q$ can be generalized into a new one, $\hQ$, by addition of terms that are linear in the functional derivatives with respect to the sources $u$ and $v$, in such a way that  
\be\label{ext}
\hQ(\Sigma[u,v])  =\Q(\Sigma) + \hQ_{|\Sigma} \Upsilon + \int d^4
x \trace \biggl( \frac{\delta^R \Upsilon}{\delta A^\mu}
\frac{\delta^L \Upsilon}{\delta A^*_\mu} + \frac{\delta^R
\Upsilon}{\delta \lambda} \frac{\delta^L \Upsilon}{\delta
\overline{\lambda}^*}+ \frac{\delta^R \Upsilon}{\delta \phi^i}
\frac{\delta^L \Upsilon}{\delta \phi^*_i} \biggr) =0
\ee
Indeed,   if we were to compute $\Q(\Sigma[u,v])$ without taking
into account the transformations of the sources $u$ and $v$, the breaking of the Slavnov--Taylor identity
would be a local functional linear in the set of gauge-invariant  local
polynomials in the physical fields, $A^*_\mu$, $c^*$,
$\phi^*_i$ and $\lambda^*$.

Eq.~(\ref{ext}) defines the 
transformations  $ \hQ_{|\Sigma}$   of   the   sources $u$ and $v$. 
Simplest examples for the transformation laws of   the $u$'s are for instance 
\bea
\hQ_{|\Sigma} u_{ij} &=& - i[\gamma^\mu \tau_{\{i} \epsilon]_\alpha \partial_\mu
u^\alpha_{j\}} + \partial_\mu \partial^\mu v_{\{ij\}} + 2 u_{\{i|k}
{v_{j\}}}^k + 2
u^\alpha_{\{i} v_{j\}\alpha} - i \partial_\mu ( u_{\{i|k} {v^\mu _{j\}}}^k
+ u^\alpha_{\{i} v^\mu_{j\}\alpha}) \CR
\hQ_{|\Sigma} u_i^\alpha &=& [\overline{\epsilon} \tau^j ]^\alpha \scal{ u_{ij} -
 i \partial_\mu ( \uk^\mu_{ij} + \ub^\mu_{ij} )} - 2i [\overline{\epsilon} \gamma_\mu
]^\alpha \partial_\nu ( \uk_i^{\mu\nu} + \ub_i^{\mu\nu} )+ i {[\gamma^\mu]_\beta}^\alpha
\partial_\mu v^\beta_i \CR
& & \hspace{10mm} - u_{ij} \vo{0}^{j\alpha} - u^\alpha_j {v_i}^j
+ u^\beta_i {v^\alpha}_\beta + u^{\alpha\beta} v_{i\beta} +
i \partial_\mu ( u_{ij} \vo{-1}^{j\mu\alpha} - u^\beta_i v_\beta^{\mu \alpha} )
\eea
These transformations are quite complicated in their most general expression. However, for
many practical computations of     non-supersymmetric 
 local counterterms%  that restore the 
% Slavnov--Taylor identity on the 1PI generating functional $\Gamma$ at
% a given order
, we can consider them at $v=0$. We  define $ Q u \equiv \scal{\hQ_{|\Sigma} u}_{|v=0}$. By using $\susy \Upsilon[u] + \Upsilon[ Q u ] = 0$ we can in fact conveniently compute $Qu$. Notice that $Q$ is not nilpotent on the sources, but we have
the result that $\Upsilon[Q^2 u]$ is a linear functional of the equation of
motion of the fermion $\lambda$.

 It is a  well-defined process to compute all
observables, provided that a complete set of sources has been
introduced. This lengthy process cannot be avoided because  there  exists no 
regulator that preserves both gauge invariance and
supersymmetry. We must keep in mind that   renormalization generally
mixes physical observables with BRST-exact operators, and a careful
analysis must be done \cite{zuber}.

\section{Enforcement of supersymmetry}

Once both Ward identities for the Green functions of fields and of observables have been established,  it  is  a straightforward (but tedious)  task to adjust the counterterms that  
  are necessary to ensure supersymmetry and gauge symmetry at the quantum level. The possibility of that is warranted by the fact the theory is renormalizable by power counting, that no anomaly exist, and that the lagrangian is stable. The technical details are given in \cite{prescription}. The question of not having a regulator that maintains  supersymmetry is irrelevant. However, in practice, 
 one wishes  to preserve the symmetry of the bare action as much as  it is  possible, and thus, one  uses dimensional reduction regularization, as in~\cite{siegel}.

\section{Twisted variables}

Using twisted variables for the spines in four dimensions allows one to extract subalgebra  of supersymmetry transformations that close  without using equations of motion \cite{beta}.  This property allows one to greatly simplify the proofs of finiteness in supersymmetric theories. Before coming to this point, let  us sketch the way the twist works for the $\N=4$ theory, by choosing the so-called first twist of this theory.

\subsection{$\mathcal{N}=4$ super-Yang--Mills theory in the twisted variables}
%\subsection{Fields and symmetries}

%Consider the $\mathcal{N}=4$ multiplet $(A,\lambda^\alpha,\phi^i)$ in a
%flat euclidean space $Spin(4) \cong SU(2)_+\times SU(2)_- $, where
%$\alpha, i$ are indices in the $\bf 4$ and the $\bf 6$ of the internal symmetry $SL(2,\mathds{H})$, which   
%is the euclidean version of the $SU(4)$ R-symmetry in Minkowski
%space. 
The components of spinor and scalar fields
$\lambda^\alpha$ and $\phi^i$ can be twisted, i.e., decomposed on
irreducible representations of the following
subgroup\footnote{Usually, one means by twist a redefinition of
 the energy momentum tensor that we do not consider here.}
\be
SU(2)_+\times {\rm diag} \scal{SU(2)_-\times SU(2)_R} \times U(1)
\subset SU(2)_+\times SU(2)_-\times SL(2,\mathds{H})
\ee
We redefine $SU(2) \cong {\rm diag}\scal{SU(2)_-\times
 SU(2)_R}$. The $\mathcal{N}=4$ multiplet is decomposed as
 follows
\be\label{dec}
( A_\mu, \Psi_\mu, \eta, \chi^I, \Phi, \bar \Phi)
\hspace {10mm}
(L, h_I, \bar \Psi_\mu, \bar \eta, \bar \chi_I)
 \ee
In this equation, the vector index $\mu$ is a ``twisted world
index'', which stands for the $(\frac{1}{2}, \frac{1}{2})$
representation of $ SU(2)_+\times SU(2)$. The index $I$ is for the
adjoint representation of the diagonal $SU(2)$. In fact, any given
field $X^I$ can be identified as a twisted antiselfdual $2$-form
$X_{\mu\nu^-}$, 
\be
X_{\mu\nu^-}\sim X_{I}
\ee
by using the flat hyperK\"{a}hler structure $J^I_{\mu\nu}$.

All  16 components of  the  $SL(2,\mathds{H})$-Majorana spinors can therefore be  mapped on the following multiplets of tensors.
 \be
 \lambda 
\to ( \Psi^\ur{1}_\mu,   \bar   \Psi^\ur{-1}_\mu, \chi^\ur{-1}_{I},     \bar  \chi^\ur{1}_{I}   ,  \eta^\ur{-1},   \bar \eta^\ur{1} )
\ee
The  scalars $\phi^i$ in the fundamental representation of $SO(6)$  decompose as follows
\be
\phi^i \to (\Phi^\ur{2},\bar \Phi^\ur{-2}, L^\ur{0}, h^\ur{0}_{I})
\ee
where the superscript states for the $U(1)$ representation. The 16 generators of the supersymmetry algebra  and the corresponding parameter $\epsilon$ are  respectively  twisted into 
\be
{\cal Q}^\ur{1},\bar {\cal Q}^\ur{-1},\, {\cal Q}^\ur{1}_\mu,\, \bar {\cal Q}^\ur{-1}_\mu, \,      {\cal Q}^\ur{1}_{I},\, \bar   {\cal Q}^\ur{-1}_{I}\ee
 and   
\be
 \epsilon 
 \to (  \omega^\ur{1},   \varpi^\ur{-1},  \varepsilon^{\ur{1} \mu},   \bar    \varepsilon ^{\ur{-1} \mu},  \upsilon^{\ur{1}  I} ,  \bar    \upsilon ^{\ur{-1} I} )
\ee
with 
\be \susy =\varpi {\cal Q}+  \omega \bar {\cal Q}+   \bar   \varepsilon ^\mu  {\cal Q}_\mu,+
   \varepsilon ^\mu \bar {\cal Q}_\mu +  \bar   \upsilon ^{I}       {\cal Q}_{I}+   \upsilon ^{I} \bar   {\cal Q}_{I} \ee

The ten-dimensional super-Yang--Mills theory determines by dimensional
reduction the untwisted $\mathcal{N}=4$ super-Yang--Mills
theory. Analogously, 
 the twisted eight-dimensional $\mathcal{N}=2$ theory determines 
the twisted formulation of the $\mathcal{N}=4$
super-Yang--Mills theory in four dimensions   by dimensional
reduction \cite{BKS, beta}.

The twisted $\mathcal{N}=2,\ d=8 $ symmetry contains     a maximal supersymmetry subalgebra
that closes without the equations of motion.   It depends on  nine twisted
supersymmetry parameters, which are one scalar $\varpi$ and one
eight-dimensional vector $\varepsilon^M$. 

By dimensional reduction $(\varpi , \varepsilon ^M  )$ decomposes into $ (
\varpi ,     \varepsilon^\mu,  \omega,   \upsilon^I  )$ and the off-shell representation of supersymmetry remains.  The dimensionally reduced four-dimensional  supersymmetry with 9 parameters  is
\be
\susy =\varpi {\cal Q}+  \omega \bar {\cal Q}+     \varepsilon ^\mu \bar {\cal Q}_\mu 
 +    \upsilon ^{I} \bar   {\cal Q}_{I} 
\ee
It closes independently of equations of motions to
\be
\susy^2 = \delta^{\rm gauge} ( \hat \Upphi(\phi) +  \varpi   \varepsilon ^\mu A_\mu)   +\varpi   \varepsilon ^\mu   \partial_\mu
\ee
with 
\be \hat \Upphi(\phi) \equiv \varpi^2 \Phi
 +\omega\varpi L
 + \varpi    {  \upsilon}^I  h_I
 +( \omega^2 +    \varepsilon^{\mu}  \varepsilon_{\mu} + {  \upsilon}^I {  \upsilon}_I)\bar\Phi
\ee

Moreover, using the extended nilpotent differential $d +\s  + Q - \varpi i_{\varepsilon} $, the action  of  $Q$ and $\s$ on all fields  is simply given by the definition of the following extended curvature
\be
\label{extcurvatur}
{\cal F} \equiv 
( d +\s  + Q - \varpi i_{\varepsilon} ) \scal{ A + \Omega + c } +
 \scal{ A + \Omega + c}^2  
  =F+\hat \Uppsi(\lambda)+ \hat \Upphi(\phi)
\ee
and the Bianchi relation that it satisfies
\begin{gather}\label{extbianchi}
 ( d + \s + Q - \varpi i_{\varepsilon} ){\cal F}+ \,[
A +\Omega + c \,,\, {\cal F}] = 0
\end{gather}
Here the linear function of the gluini $\hat \Uppsi(\lambda)$ is\footnote{Given a vector field $V$, one defines the $1$-form  $g(V ) \equiv g_{\mu\nu} V^\mu dx^\nu$, and the vector  $(J_I (V))^\mu\equiv {{J_I}^\mu}_\nu    V^\nu$. $\hat \Uppsi(\lambda)_\mu $ can be written $\varpi \Psi_\mu + 
\omega \bar \Psi_\mu  +{  \upsilon}_{\mu\nu^-}  {\bar \Psi} ^{\nu} -{  \varepsilon_{\mu}}   \eta +  
 { \varepsilon} ^{\nu}  \chi_{\mu\nu^-} $}

\be
\hat \Uppsi(\lambda)\equiv \varpi \Psi+  \omega \bar \Psi +{  \upsilon}^I J_I( {\bar \Psi})   +  g(\varepsilon) \eta + i_\varepsilon \chi 
\ee
Eqs.~(\ref{extcurvatur}) and (\ref{extbianchi})  determine respectively  the action of $Q$ and $\s$ on $A$, $c$, $\Omega$ and on  the fields on the right-hand-side of Eq.~(\ref{extcurvatur}), by expansion in form degree.

%Eqs.~(\ref{extcurvatur}) and (\ref{extbianchi})  indicate that  the $\s$ transformations of physical fields are their gauge
%transformations with parameter $\Omega$, and that the 
%  $Q$ transformations  of   
%physical fields decomposes as a gauge transformations  with 
%parameter $c$ and a supersymmetry transformation     $\susy$ with parameters 
%$\omega,\varpi,{\varepsilon}, \upsilon_I$,  $Q= \susy -\delta^{\rm gauge }(c) $.
%One has in fact
%\be
%( d +\s  + Q - \varpi i_{\varepsilon} )^2=0
%\ee
%by construction, and thus 
%\be (\susy)^2 = \delta^{\rm gauge}({\cal M}(\phi) + \varpi
%i_{\varepsilon} A) + \varpi \L_{\varepsilon} \label{closed}\ee
%with  $ Q c = {\cal M}(\phi) + \varpi i_{\varepsilon} A - c^2 $ and 
%\be
%{\cal M}(\phi)\equiv \varpi^2 \Phi + \varpi \omega L + \varpi
%\upsilon_I h^I + (\omega^2 + \upsilon_I \upsilon^I + |{\varepsilon}|^2
%)\bar \Phi
%\ee
% This is equivalent to $s^2=sQ+Qs=0$ and  $Q^2= \varpi  \L_ \epsilon $. In fact, Eqs.~(\ref{extcurvatur}) and (\ref{extbianchi}) justify the introduction of shadows, from a geometrical point of view. 

Few degenerate component equations  occur when solving Eqs.~(\ref{extcurvatur}) and (\ref{extbianchi}). They are solved by introducing the fields     $\bar \chi_I$ and $\bar \eta$, the  auxiliary fields $H^I,T_\mu$ and the shadow field $\mu$. Notice that the auxiliary fields $H_I$ and $T_\mu$, carry a
total of $7=3+4$ degrees of freedom. The  latter compensate the  deficit between the number of off-shell gauge-invariant degrees of freedom of fermions and bosons in the theory.

Eqs.~(\ref{extcurvatur}) and (\ref{extbianchi}) determine $\susy$ as 
\bea
\label{susy}
 \susy A &=& \varpi \Psi + \omega \bar \Psi + g({\varepsilon})
\eta + g(J_I {\varepsilon} ) \chi^I + \upsilon_I J^I (\bar \Psi)\CR
\susy \Psi &=& - \varpi d_A \Phi - \omega \scal{ d_A L + T}+
i_{\varepsilon} F + g(J_I {\varepsilon}) H^I + g({\varepsilon}) [\Phi,
\bar \Phi] - \upsilon_I \scal{ d_A h^I + J^I (T)}\CR \susy \Phi
&=& - \omega \bar \eta + i_{\varepsilon} \Psi - \upsilon_I \bar
\chi^I \CR
\susy \bar \Phi &=& \varpi \eta \CR
\susy \eta &=& \varpi [\Phi, \bar \Phi] - \omega [\bar \Phi, L] +
\Lc_{\varepsilon} \bar \Phi - \upsilon_I [\bar \Phi, h^I]\CR
\susy \chi^I &=& \varpi H^I + \omega [\bar \Phi, h^I] + \Lc_{J^I
 {\varepsilon}} \bar \Phi- \upsilon_I [\bar \Phi, L] +
{{\varepsilon}^I}_{JK} \upsilon^J [\bar \Phi, h^K] \CR
\susy H^I &=& \varpi [\Phi, \chi^I] + \omega \scal{ [L, \chi^I] -
 [\eta, h^I] - [\bar \Phi, \bar \chi^I]} - \Lc_{J^I
 {\varepsilon}} \eta - [\bar \Phi, i_{J^I {\varepsilon}} \Psi] +
\Lc_{\varepsilon} \chi^I\CR
& & \hspace{15mm} + \upsilon_J [h^J , \chi^I] + \upsilon^I \scal{
 [\eta, L] + [\bar \Phi, \bar \eta]} - {{\varepsilon}^I}_{JK}
\upsilon^J \scal{[\eta, h^K] + [\bar\Phi, \bar\chi^K]} \CR
\susy L &=& \varpi \bar \eta - \omega \eta + i_{\varepsilon} \bar
\Psi - \upsilon_I \chi^I\CR
\susy \bar \eta &=& \varpi [\Phi, L] + \omega [\Phi, \bar \Phi] +
\Lc_{\varepsilon} L + i_{\varepsilon} T + \upsilon_I \scal{ H^I + [h^I, L]}\CR
\susy \bar \Psi &=& \varpi T - \omega d_A \bar \Phi -
g({\varepsilon}) [\bar \Phi, L] + g(J_I {\varepsilon}) [ \bar \Phi, h^I]
+ \upsilon_I J^I (d_A \bar \Phi)\CR
\susy T &=& \varpi [\Phi, \bar \Psi] + \omega \scal{-d_A \eta -
 [\bar\Phi, \Psi] + [L, \bar \Psi]} - g({\varepsilon})\scal{[\eta, L] +
 [\bar\Phi, \bar \eta]} \CR
& & \hspace{15mm} + g(J_I {\varepsilon}) \scal{ [\eta, h^I] +
 [\bar\Phi, \bar \chi^I]} + \Lc_{\varepsilon} \bar \Psi + \upsilon_I \scal{
 [h^I, \bar \Psi] + J^I (d_A \eta + [\bar\Phi, \bar \Psi])}\CR
\susy h^I &=& \varpi \bar \chi^I + \omega \chi^I - i_{J^I
 {\varepsilon}} \bar \Psi - \upsilon^I \eta - {{\varepsilon}^I}_{JK}
\upsilon^J \chi^K \\*
\susy \bar \chi^I &=& \varpi [\Phi, h^I] + \omega \scal{[L, h^I]
 - H^I} + \Lc_{\varepsilon} h^I - i_{J^I {\varepsilon}} T + \upsilon^I
[\Phi, \bar \Phi] + \upsilon_J [h^J, h^I] + {{\varepsilon}^I}_{JK}
\upsilon^J H^K \nonumber
\eea
 One can verify that, for $T_\mu=H_I=0$, the transformation laws of $\susy$  in Eq.~(\ref{susy}) are the on-shell transformation laws of the twisted $\N=4$ supersymmetry. It is quite remarkable that the supersymmetry transformations are the solution of the curvature equation~(\ref{extcurvatur}) and its Bianchi identity (\ref{extbianchi}). As we will shortly sketch, these equations play a key role in  non-renormalization theorems.

\subsection{Protected operators}
Superconformal invariance  implies  that  the so-called BPS local operators are  protected from renormalization and their anomalous dimensions vanish  to all orders in perturbation theory \cite{BPS}. In the $\N=4$ theory,  these operators  play an important role for  the $AdS/CFT$ correspondence, since their non-renormalization properties allows  to test the conjecture.  

One wishes  to prove that,  without the assumption of the superconformal symmetry, $\N=4$ supersymmetry implies that all $1/2$ BPS primary operators, and thus all their descendants, have zero anomalous dimension. We will sketch the proof of this statement using only Ward identities associated to gauge and supersymmetry invariance. The $1/2$ BPS primary operators are  the gauge-invariant polynomials in the scalar fields of the theory in traceless symmetric representations of the $SO(5,1)$ R-symmetry group.

In the gauge $\varepsilon^\mu = 0$ the operator $Q$ is nilpotent.\footnote{Remember that the supersymmetry parameters appearing in the differential $Q$ can be understood in quantum field theory as  gauge parameters of the $Q$-invariant gauge-fixing action.} The linear function of the scalar fields $\hat \Upphi(\phi)$  that characterizes the field dependent gauge transformations that appear in the commutators of two supersymmetries,
depends in this case on five parameters, \be \hat \Upphi(\phi) =
\varpi^2 \Phi + \varpi \omega L + \varpi \upsilon_I h^I +
(\omega^2 + \upsilon_I \upsilon^I )\bar \Phi \ee

The decomposition under the independent functions of the supersymmetric parameters of the invariant polynomial $\mathcal{P}$ in $\hat \Upphi(\phi)$ gives  all the  gauge invariant polynomials in the scalar fields that belongs to traceless symmetric representations of $SO(5,1)$ \cite{beta}. Since  $Q$ is nilpotent with the  restricted set of parameters,  the shadow number $2$ component of the  curvature equation  (\ref{extcurvatur}) is also a curvature equation
\be Q      c     + c^2   = \hat \Upphi(\phi) \ee
By comparison with the  Baulieu--Singer curvature equation in TQFT's, one interprets $c$ as the component of the connexion of the space of gauge orbits along the fundamental vector field generating supersymmetry and   $ \hat \Upphi(\phi)$ as the component of its curvature along the same fundamental vector field.\footnote{By this we mean the following. Given $\omega$ as the connection of the fiber bundle defined as the direct sum of the space of irreducible connexions and the space of matter fields of the theory, on which the group of pointed gauge transformations acts freely.  Define $\Phi$ as the corresponding  curvature $s\,\omega + \omega^2$. The supersymmetry transformations can be seen as generated by an anticommuting fundamental vector field $v$, such that $ Q = L_v \equiv  [ I_v , s]$. With the reduced set of parameters, the vector field $v$ commutes with itself. Then one has
\be L_v  I_v \omega + ( I_v \omega )^2 = \frac{1}{2} {I_v }^2 \scal{ s\,  \omega + \omega ^2} + \frac{1}{2} [L_v, I_v] \omega =   \frac{1}{2} {I_v }^2 \Phi\nonumber \ee}  The Chern--Simons formula then implies that any given invariant polynomial $\mathcal{P}(\hat \Upphi)$ can be written as a $Q$-exact term
\be \mathcal{P}\scal{\hat \Upphi (\phi)} = Q \,\Delta\scal{c,\hat \Upphi(\phi)} \ee
where the Chern-Simons form $\Delta$ is given  by 
\be \Delta\scal{c, \omega(\varphi)} \equiv \int_0^1 dt\,
\mathcal{P}\scal{c\,|\, t \omega(\varphi) + (t^2- t) c^2\,} \ee  

Any given polynomial in the scalar fields belonging to a traceless
symmetric representation of $SO(5,1)$ has a canonical dimension
which is strictly lower than that of all other
operators in the same representation, made out of
other fields. Thus, by
 power counting, the polynomials in the scalar fields can
 only mix between themselves under renormalization. Thus, if 
 \def\C{{\cal C}}
 $\C$ is the Callan--Simanzik operator,   for any homogeneous polynomial $\mathcal{P}_A$ of degree $n$ in
the traceless symmetric representation, renormalization can only produce anomalous dimensions that satisfy \be \C \bigl[
\mathcal{P}_A\scal{\hat \Upphi(\phi)} \cdot \Gamma \bigr] = \sum_B
{\gamma_A}^B \bigl[ \mathcal{P}_B \scal{\hat \Upphi(\phi)} \cdot
\Gamma \bigr] \ee In this  notation, given a local operator $\cal O$,   $ \bigl[{\cal O } \cdot
\Gamma \bigr]$ means its  insertion in the   generating functional  of one-particle irreducible Green functions $\Gamma$.  Then, the Slavnov--Taylor identities imply \be
\C \bigl[ \Delta_A\scal{c,\hat \Upphi(\phi)} \cdot \Gamma \bigr] =
\sum_B {\gamma_A}^B \bigl[ \Delta_B\scal{c,\hat \Upphi(\phi)} \cdot
\Gamma \bigr] + \cdots \ee
where the dots stand for possible
$\Q_{|\Gamma}$-invariant corrections. However, in the shadow-Landau
gauge (i.e., the gauge (\ref{SLG}) with $\alpha=0$), $\Delta_A(c,\hat \Upphi(\phi))$ cannot appear in the right-hand-side because such term would break the so-called ghost Ward
identities \cite{beta}. One thus gets the result that ${\gamma_A}^B = 0$
\be \C \bigl[ \mathcal{P}_A\scal{\hat \Upphi(\phi)} \cdot \Gamma \bigr] = 0 \ee
 Upon decomposition of this equation in function of the five
independent supersymmetry parameters, one then gets the finiteness
proof for each invariant polynomial
$\mathcal{P}(\phi)\equiv \mathcal{P}(\phi^i,\phi^j,\phi^k,\cdots)$
in the traceless symmetric representation of the R-symmetry group,
 namely \be \C \bigl[ \mathcal{P}(\phi) \cdot \Gamma
\bigr] = 0 \label{BPS}\ee

Having proved that all $1/2$ BPS
primary operators have zero anomalous dimension,
the $Q$-symmetry implies that
 all the operators generated from them, by applying 
 $\mathcal{N}=4$ super-Poincar\'e generators, have also vanishing
 anomalous dimensions. It follows that all the operators of
the $1/2$ BPS multiplets are protected operators.

It is worth considering as an example the simplest case of
$\trace \hat \Upphi(\phi)^2$. One has
\begin{gather}
Q \trace \scal{\hat \Upphi(\phi) c - \frac{1}{3} c^3} = \trace
\hat \Upphi(\phi)^2 \hspace{10mm} \s Q \trace \scal{\hat \Upphi(\phi) c -
 \frac{1}{3} c^3} = 0 \CR
\s \trace \scal{\hat \Upphi(\phi) c - \frac{1}{3} c^3} = \trace \Scal{
 \mu\scal{ \hat \Upphi(\phi) - c^2 } - [\Omega, \hat \Upphi(\phi)] c}
\end{gather}

These constraints imply that 
 $\Delta^\gra{0}{3}_{[\frac{3}{2}]}$ is proportional to $\trace
 \scal{\hat \Upphi(\phi) c - \frac{1}{3} c^3}$. Thus the three
 insertions that we have introduced can only be multiplicatively
 renormalized, with the same anomalous dimension.
 Moreover, the ghost Ward identities forbid the introduction of
 any invariant counterterm depending on  the shadow field $c$, if it is
 not trough a derivative term $d c$ or particular combinations of
 $c$ and the other fields that do not appear in the insertion
 $\trace \scal{\hat \Upphi(\phi) c - \frac{1}{3} c^3}$.
 This gives the result that \be \C\, \bigl[ \trace  \hat \Upphi(\phi)^2 \, \cdot
\Gamma \bigr] = 0 \ee

Finally, by  decomposition  of  the   gauge-invariant operators   upon independent combinations of the    parameters, we obtain that all the 20 operators that constitute the traceless-symmetric tensor
representation of rank two in $SO(5,1)$ are protected operators
\begin{gather}
\trace\scal{ \Phi^2}\, , \ \trace\scal{ \Phi L} \, , \
\trace\scal{\Phi \bar \Phi + \frac{1}{2} L^2}\, ,\ \trace \scal{\bar \Phi L}\,
,\ \trace \scal{\bar \Phi^2 }\,,\label{sst}\CR
\trace\scal{ \Phi h_I} \, , \ \trace\scal{ L h_I }\, , \ \trace\scal{
 \bar\Phi h_I} \, , \ \trace\scal{ \delta_{IJ} \Phi \bar \Phi +
 \frac{1}{2} h_I h_J} \label{protected}
\end{gather}
This constitutes the simplest application  of Eq.~(\ref{BPS}), for $\mathcal{P}(\phi)
\equiv \trace \scal{ \phi^i \phi_j -\frac{1}{6}
 \delta^{i}_{j} \phi_k \phi^k }$.

\subsection{Cancellation of the $\beta$ function form descent equations}
To show  that the coupling constant  of the $\N=4$ theory is not rescaled by renormalization, the key point  is  proving that the action $ S=\int \L_4^0$  has vanishing anomalous dimension,  in the sense that it cannot   be renormalized by  anything but a  mixing with a BRST-exact counterterms. We will restrict here to the proof of this lemma, that is   proving the Callan--Symanzik equation
\bea \label{zeroZ}
  \C \ \bigl[\int \L_4^0 \cdot
 \Gamma \bigr] = \ins{1} 
 \eea where $\hat \Uppsi^\ur{1} $ is a functional of ghost number -1 and shadow number 0. (See \cite{beta} for a complete discussion.)

 To prove (\ref{zeroZ}), we will use the fact that  descent equations imply that  the
lagrangian density is uniquely linked to  a    combination of    protected operators (\ref{protected}), with coefficients that are  fixed functions of the supersymmetric parameters. 

As shown in \cite{beta},  the reduced supersymmetry with the  six generator ${\cal Q},\, \bar {\cal Q}$ and $\bar{\cal Q}_\mu$  is sufficient to  completely   determine the classical action. For simplicity, we will thus restrict $\susy$ to these generators in this section ($\upsilon^I = 0$).  Because $\L^0_4 $ and $Ch^0_4=Tr (FF) $  are supersymmetric invariant only modulo a boundary-term, the
algebraic Poincar\'e lemma predicts series of cocycles, which are   linked to $\L^0_4 $ and $Ch^0_4 $ by descent equations, as follows: 
\be\begin{split}
\susy \L^0_4 + d \L^1_3 &= 0 \\*
\susy \L^1_3 + d \L^2_2 &= \varpi i_\varepsilon \L^0_4 \\*
\susy \L^2_2 + d \L^3_1 &= \varpi i_\varepsilon \L^1_3 \\*
\susy \L^3_1 + d \L^4_0 &= \varpi i_\varepsilon \L^2_2 \\*
\susy \L^4_0 &= \varpi i_\varepsilon \L^3_1
\end{split}\hspace{20mm}\label{cocy}\begin{split}
\susy Ch^0_4 + d Ch^1_3 &= 0 \\*
\susy Ch^1_3 + d Ch^2_2 &= \varpi i_\varepsilon Ch^0_4 \\*
\susy Ch^2_2 + d Ch^3_1 &= \varpi i_\varepsilon Ch^1_3 \\*
\susy Ch^3_1 + d Ch^4_0 &= \varpi i_\varepsilon Ch^2_2 \\*
\susy Ch^4_0 &= \varpi i_\varepsilon Ch^3_1
\end{split}\ee
Using the grading properties of the shadow number and the form degree,
we conveniently define
\begin{gather}\L \equiv \L_4^0 + \L_3^1 + \L_2^2 + \L_1^3 + \L_0^4\CR
 Ch  \equiv  Ch_4^0 + Ch_3^1 + Ch_2^2 + Ch_1^3 + Ch_0^4 \end{gather}
The descent equations can then be written in a unified way \be (d +
\susy - \varpi i_\varepsilon ) \L = 0 \hspace{10mm}(d +
\susy - \varpi i_\varepsilon ) Ch = 0 \ee Note that on
gauge-invariant polynomials in the physical fields, $\susy$ can be
identified to $\s + Q$, in such way that the differential $(d +
\susy - \varpi i_\varepsilon )$ is nilpotent on them. Since
$\L^0_4 $ and $Ch^0_4 $ are the unique solutions of the first
equation in (\ref{cocy}), one obtains that $\L$ and
$Ch$ are the only non-trivial solutions of the descent equations,
that is, the only ones that cannot be written as
$(d+\susy-\varpi i_\varepsilon)\, \Xi$ for a non trivial element of the $\s$ cohomology $\Xi$.  The expression of the cocycles $Ch^{s}_{4-s}$ can be simply obtained using the extended curvature (\ref{extcurvatur}) since the extended second Chern class 
\be Ch = \frac{1}{2} \trace \Bigl( F + \varpi \Psi + \omega \bar \Psi
+g(\varepsilon) \eta + g(J_I \varepsilon) \chi^I + \varpi^2 \Phi +
\varpi \omega L + (\omega^2 + |\varepsilon|^2) \bar\Phi \Bigr)^2 \ee
is $(d+\susy-\varpi i_\varepsilon)$ invariant by definition.

As for determining the explicit form of $\L^{s}_{4-s}$ for $s
\geqslant 1$, we found no other way than doing a brute force
computation. In this way, one gets   \cite {beta}
\be \L^4_0 = \frac{1}{2} \trace \Scal{\scal{ \varpi^2
 \Phi + \varpi \omega L + \omega^2 \bar \Phi}^2 +
 \varpi^2 |\varepsilon|^2 \bar \Phi^2 } \ee

 The last cocycle $\L^4_0 $ is a linear combination of the protected operators (\ref{protected}) and thus, its anomalous dimension is zero. This permits to prove that its ascendant $\L^0_4 $ can only be renormalized by $d$-exact or $\S_{|\Sigma}$-exact counterterms.
  
 \section{Conclusion}  
In the formalism that we have presented,  the set of fields  of a supersymmetric theory  has been extended.   With the introduction of shadow fields, one  can express supersymmetry under the form of a nilpotent differential operator.

This clarifies many questions that arise when one builds the quantum field theory of a supersymmetric Yang--Mills theory, in particular for defining observables and study their renormalization.  For instance, 
 supersymmetric observables can  be defined within the standard point of view of   the cohomology of the BRST symmetry. In this framework,  we have been able to define unambiguously the computation at all order in perturbation theory of all correlation functions,  including insertions  of  gauge invariant local operators. The Slavnov--Taylor identities permit one to compute the non-invariant finite counterterms to maintain supersymmetry and gauge invariance of observables, independently of the choice   of  the regularization scheme.

By  twisting the spinors, one can  find subalgebra of  supersymmetry with no equations of motions in the closure relations. This permits to simplify  the proofs of various renormalization theorems for the $\N=4$ super-Yang--Mills theory.

\subsection*{Acknowledgments}

This work was partially supported under the contract ANR(CNRS-USAR) \\ \texttt{no.05-BLAN-0079-01}.


\begin{thebibliography}{99}

\bibitem{shadow}
  L.~Baulieu, G.~Bossard and S.~P.~Sorella,
  ``Shadow fields and local supersymmetric gauges,''
  Nucl.\ Phys.\ B {\bf 753}, 273 (2006)
  \texttt{[arXiv:hep-th/0603248]}.
  %%CITATION = HEP-TH 0603248;%%

\bibitem{stockinger}
  W.~Hollik, E.~Kraus and D.~St\"{o}ckinger,
  ``Renormalization and symmetry conditions in supersymmetric QED,''
  Eur.\ Phys.\ J.\ C {\bf 11}, 365 (1999)
  \texttt{[arXiv:hep-ph/9907393]};
  %%CITATION = HEP-PH 9907393;%%
  W.~Hollik and D.~St\"{o}ckinger,
  ``Regularization and supersymmetry-restoring counterterms in supersymmetric
  QCD,''
  Eur.\ Phys.\ J.\ C {\bf 20}, 105 (2001)
  \texttt{[arXiv:hep-ph/0103009]};
  %%CITATION = HEP-PH 0103009;%%\\
  I.~Fischer, W.~Hollik, M.~Roth and D.~St\"{o}ckinger,
  ``Restoration of supersymmetric Slavnov--Taylor and Ward identities in
  presence of soft and spontaneous symmetry breaking,''
  Phys.\ Rev.\ D {\bf 69}, 015004 (2004)
  \texttt{[arXiv:hep-ph/0310191]}.
  %%CITATION = HEP-PH 0310191;%%


\bibitem{beta}
 L.~Baulieu and  G.~Bossard,
  `New Results on $\N=4$ Super-Yang--Mills Theory,'' 
  Phys.\ Lett.\ B {\bf 632}, 131 (2006),
  \texttt{[arXiv:hep-th/0507003]}; 
  L.~Baulieu, G.~Bossard and S.~P.~Sorella,
  ``Finiteness properties of the $\N = 4$ super-Yang--Mills theory in
  supersymmetric gauge,'' 
  Nucl.\ Phys.\ B {\bf 753}, 252 (2006)
  \texttt{[arXiv:hep-th/0605164]}.
  
  
\bibitem{henneaux}
  M.~Henneaux,
  ``Remarks on the renormalization of gauge-invariant  operators in Yang--Mills
  theory,''
  Phys.\ Lett.\ B {\bf 313}, 35 (1993)
  \texttt{[arXiv:hep-th/9306101]}.
  %%CITATION = HEP-TH 9306101;%%

\bibitem{zuber}
   H.~Kluberg--Stern and J.~B.~Zuber,
   ``Ward identities and some clues to the renormalization of gauge
  -invariant  operators,''
   Phys.\ Rev.\ D {\bf 12}, 467 (1975);
   %%CITATION = PHRVA,D12,467;%%
   H.~Kluberg--Stern and J.~B.~Zuber,
   ``Renormalization of nonabelian gauge theories in a background
   field gauge. 2. Gauge-invariant  operators,'' 
   Phys.\ Rev.\ D {\bf 12}, 3159 (1975);
   %%CITATION = PHRVA,D12,3159;%%
   S.~D.~Joglekar and B.~W.~Lee,
   ``General theory of renormalization of gauge-invariant  operators,''
   Annals Phys.\  {\bf 97}, 160 (1976).
   %%CITATION = APNYA,97,160;%%

\bibitem{prescription}
    L.~Baulieu and  G.~Bossard,
``Supersymmetric renormalization prescription in $\N = 4$ super-Yang--Mills theory,"
    Phys. Lett.  B643 (2006) 294-302,  
      \texttt{[arXiv:hep-th/0609189]}.

\bibitem{siegel}
  W.~Siegel,
  ``Supersymmetric dimensional regularization via dimensional reduction,''
  Phys.\ Lett.\ B {\bf 84}, 193 (1979);\\
 W.~Siegel,
  ``Inconsistency of supersymmetric dimensional regularization,''
  Phys.\ Lett.\ B {\bf 94}, 37 (1980);\\
  %%CITATION = PHLTA,B94,37;%%
  L.~V.~Avdeev, S.~G.~Gorishnii, A.~Y.~Kamenshchik and S.~A.~Larin,
  ``Four loop beta function in the Wess--Zumino model,''
  Phys.\ Lett.\ B {\bf 117}, 321 (1982).
  %%CITATION = PHLTA,B117,321;%%




\bibitem{BKS}
  L.~Baulieu, H.~Kanno and I.~M.~Singer,
  ``Special quantum field theories in eight and other dimensions,''
  Commun.\ Math.\ Phys.\  {\bf 194} (1998) 149
  \texttt{[arXiv:hep-th/9704167]};
   L.~Baulieu, G.~Bossard and A.~Tanzini,
 ``Topological Vector Symmetry of BRSTQFT and Construction of Maximal
 Supersymmetry,''
 JHEP {\bf 0508} (2005) 037
 \texttt{[arXiv:hep-th/0504224]}.
  

  \bibitem{BPS}
    S.~Minwalla,
  ``Restrictions imposed by superconformal invariance on quantum field
  theories,''
  Adv.\ Theor.\ Math.\ Phys.\  {\bf 2}, 781 (1998)
  \texttt{[arXiv:hep-th/9712074]};
  %%CITATION = HEP-TH 9712074;%%
  J.~Rasmussen,
  ``Comments on $\N = 4$ superconformal algebras,''
  Nucl.\ Phys.\ B {\bf 593} (2001) 634
  \texttt{[arXiv:hep-th/0003035]}.
  %%CITATION = HEP-TH 0003035;%%




 
\end{thebibliography}
\end{document}